\documentclass[aps,pre,reprint,superscriptaddress,twocolumn,showkeys,amsmath,amssymb,longbibliography,floatfix,tikz]{revtex4-2}
\usepackage[english]{babel}
\usepackage{amsmath,amssymb,bbm,mathrsfs,bm,braket,color,graphicx,comment,amsfonts,dsfont}
\usepackage{hyperref}
\hypersetup{colorlinks=true,urlcolor=blue,linkcolor=blue,citecolor=blue}
\usepackage[mathscr]{euscript}
\usepackage{bm}
\usepackage{physics}
\usepackage{siunitx}

\begin{document}

\title{Non-unitary Time Evolution via the Chebyshev Expansion Method}

\author{\'{A}ron Holl\'{o}}
\affiliation{Department of Physics of Complex Systems, ELTE E\"otv\"os Lor\'and University, H 1117, Budapest, Hungary}
\affiliation{Wigner Research Centre for Physics, H-1525, Budapest, Hungary}

\author{D\'{a}niel Varjas}
\affiliation{\mbox{Institute for Theoretical Solid State Physics, IFW Dresden, 01069 Dresden, Germany}}
\affiliation{\mbox{W\"urzburg-Dresden Cluster of Excellence ct.qmat, Germany}}
\affiliation{Department of Theoretical Physics, Institute of Physics,
Budapest University of Technology and Economics, M\H{u}egyetem rkp. 3., 1111 Budapest, Hungary}

\author{Cosma Fulga}
\affiliation{\mbox{Institute for Theoretical Solid State Physics, IFW Dresden, 01069 Dresden, Germany}}
\affiliation{\mbox{W\"urzburg-Dresden Cluster of Excellence ct.qmat, Germany}}

\author{L\'{a}szl\'{o} Oroszl\'{a}ny}
\affiliation{Department of Physics of Complex Systems, ELTE E\"otv\"os Lor\'and University, H 1117, Budapest, Hungary}
\affiliation{Wigner Research Centre for Physics, H-1525, Budapest, Hungary}

\author{Viktor K{\"o}nye}
\affiliation{\mbox{Institute for Theoretical Solid State Physics, IFW Dresden, 01069 Dresden, Germany}}
\affiliation{\mbox{W\"urzburg-Dresden Cluster of Excellence ct.qmat, Germany}}
\affiliation{Institute for Theoretical Physics Amsterdam, University of Amsterdam, Science Park904, 1098 XH Amsterdam, The Netherlands}

\date{\today}
\begin{abstract}
The Chebyshev expansion method is a well-established technique for computing the time evolution of quantum states, particularly in Hermitian systems with a bounded spectrum.
Here, we show that the applicability of the Chebyshev expansion method extends well beyond this constraint: It remains valid across the entire complex plane and is thus suitable for arbitrary non-Hermitian matrices.
We identify that numerical rounding errors are the primary source of errors encountered when applying the method outside the conventional spectral bounds, and they are not caused by fundamental limitations.
By carefully selecting the spectral radius and the time step, we show how these errors can be effectively suppressed, enabling accurate time evolution calculations in non-Hermitian systems.
We derive an analytic upper bound for the rounding error, which serves as a practical guideline for selecting time steps in numerical simulations.
As an application, we illustrate the performance of the method by computing the time evolution of wave packets in the Hatano-Nelson model.
\end{abstract}

\maketitle

\section{Introduction}

Solving unitary time evolution is a fundamental challenge in quantum mechanics, essential for understanding the dynamics of quantum systems.
Although the problem is formally solved, numerically obtaining the time evolved wavefunction can run into complications, especially for large systems.
Many numerical methods and tricks have been devised to deal with this problem and optimize the calculations \cite{moler2003, moore2011}.

A frequently used method is the Chebyshev expansion method \cite{tal-ezer1984, kosloff1994a, chen1999, weisse2006a, weisse2008, wang2017}. This relies on the expansion of the exponential function using Chebyshev polynomials \cite{szego1975, mason2002}.

In recent years there has been growing interest in studying systems governed by a non-Hermitian Hamiltonian \cite{ashida2020}.
An effective non-Hermitian description can appear in many different contexts, for example the quantum dynamics of open systems, or classical dynamics of dissipative non-reciprocal systems can be described via a non-Hermitian formalism.
All of these lead to the same type of non-unitary dynamics, that can be solved in the same way, by exponentiating a matrix.

Non-Hermitian matrices often lead to numerical instabilities, due to their sensitivity to small perturbations and the absence of Weyl's inequality \cite{holbrook1992}.
More instability requires more efficient numerical methods in order to obtain accurate results, especially for larger systems.

Generalizing the Hermitian Chebyshev expansion to non-Hermitian matrices seems like a natural step; however, this ran into some complications.
According to the physics literature the Chebyshev expansion can not be used outside of the real $[-1,1]$ interval \cite{kosloff1994a, fehske2009, hatano2016a, weisse2006a, weisse2008, diogosoares2024}, where the Chebyshev polynomials are well-behaved.

To overcome this problem, many works in the physics literature introduce alternative methods to compute the time evolution of a non-Hermitian system.
These methods include algorithms based on Taylor expansion \cite{spring2024}, Runge-Kutta method \cite{noronha2022}, the usage of more general polynomials such as the Faber polynomials \cite{diogosoares2024}, and methods based on Hermitizing the Hamiltonian \cite{hatano2016a, chen2023a}.

In contrast, if one looks at the mathematics literature, it becomes evident that the Chebyshev expansion of the exponential function should work for arbitrary complex numbers \cite{szego1975, mason2002, costin2018, munch2019a}.

In this paper we show that the Chebyshev expansion method does also work for non-Hermitian Hamiltonians and can be used to numerically compute the non-unitary time evolution of arbitrary states.
We explore the apparent inconsistency between the physics and mathematics literature, and we discuss the numerical limitations of the method.

First, we will explain the exponentiation of a simple complex number (Sec.~\ref{sec:expfunc}), and then we will discuss the exponentiation of a non-Hermitian matrix (Sec.\ref{sec:expmat}) and calculate the time evolution of wave packets in a simple fruit-fly system: the Hatano-Nelson model \cite{hatano1996}.

\section{Complex exponential function}
\label{sec:expfunc}

In this section, we consider the numerical evaluation of the function $\mathrm{e}^{-itz}$ where $t>0\in\mathbb{R}$ and $z\in\mathbb{C}$ using the Chebyshev expansion.

\subsection{Chebyshev expansion of the exponential function}

Throughout this paper we will be using the Chebyshev polynomials of the first kind, $T_m(z)$, defined through the following recursion relation \cite{mason2002}:
\begin{equation}
\begin{split}
            T_0(z)&=1,\\
        T_1(z)&=z, \\
        T_{m+1}(z)&=2zT_m(z)-T_{m-1}(z).
\end{split}
\end{equation}
With this definition the Chebyshev polynomials can be defined on the entire complex plane $z\in\mathbb{C}$.

The exponential function can be expressed as a series as \cite{chen1999}
\begin{equation}
\label{eq:expansion}
        \mathrm{e}^{-itz} = J_0(t)+2\sum\limits_{m=1}^\infty (-i)^mJ_m(t)T_m(z),
\end{equation}
where $J_m$ are the Bessel functions of the first kind.
This series is convergent and valid for the entire complex plane since the exponential function is analytic on the entire plane. See, e.g., Theorem 9.1.1 in Ref.~\cite{szego1975} or Theorem 1 in \cite{costin2018} (the Chebyshev polynomials are special cases of the Jacobi polynomials).

Often times, in practical numerical implementations, the values of $z$ are restricted to the $[-1,1]$ interval on the real axis.
In the following section we will see why this is the case and how we can move away from the $[-1,1]$ interval and still obtain numerically precise results.

\subsection{Numerical accuracy}

We implemented the series expansion of the exponential function in Eq.~\eqref{eq:expansion} using a \texttt{Python} code (see Appendix \ref{sec:implementation} for details).
To assess the accuracy of our numerical values for the exponential function, we compare them with the exponential function computed using standard a implementation \cite{cody1980software} (see Appendix \ref{sec:implementation} for details).

\begin{figure}[t]
        \includegraphics[width=\columnwidth]{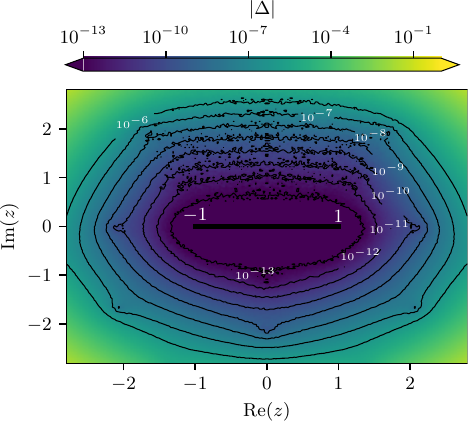} 
        \caption{\label{fig:ellipses} The absolute difference ($\Delta$) between $\mathrm{e}^{-itz}$ computed using the standard exponential function and obtained using the Chebyshev series expansion Eq.~\eqref{eq:expansion}, truncated by keeping only $m = 250$ terms in the sum, for different values of $z\in\mathbb{C}$. The results are shown for $t=8$.}
\end{figure}

The difference between our Chebyshev series implementation and that of the standard exponential function computed using floating-point arithmetic is shown in Fig.~\ref{fig:ellipses}.
As we can see, the best accuracy is achieved on the $z\in[-1,1]$ interval.
Around this region, the accuracy remains roughly constant along ellipses with focal points at $-1$ and $1$, but it decreases exponentially for larger ellipses.
How is this possible if the Eq.~\eqref{eq:expansion} expansion is valid on the entire complex plane?

\begin{figure}[t]
       \includegraphics[width=0.45\textwidth]{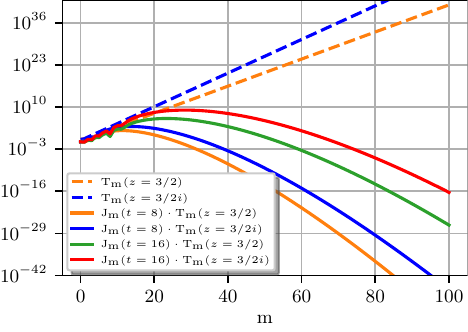}
        \caption{\label{fig:terms} Individual terms in the Chebyshev expansion \eqref{eq:expansion} of $\mathrm{e}^{-itz}$ for various values of $t={8,16}$ and $z=\{3/2,3/2i\}$ as a function of the expansion order $m$.}

\end{figure}

In order to understand this, we look at the individual terms in the expansion.
These terms are shown in Fig.~\ref{fig:terms} as a function of the order of the expansion ($m$).
Chebyshev polynomials $T_m(z)$ evaluated on the real $[-1,1]$ interval are bounded by $-1 \leq T_m(z)\leq 1$ for all values of $m$.
Away from this interval we can give bounds using the so-called Bernstein ellipses with radius $\varrho\geq 1$, defined as
\begin{equation}
    \mathcal{E}_\varrho=\left \{ z\in\mathbb{C} \bigg| z =\frac{1}{2}\left(\varrho\mathrm{e}^{i\vartheta}+\frac{1}{\varrho}\mathrm{e}^{-i\vartheta}\right),\vartheta\in[0,2\pi)\right\}.
\end{equation}
These ellipses have foci at $z=\pm1$.
For $w\in\mathcal{E}_\varrho$ the Chebyshev polynomials are bounded as \cite{mason2002}
\begin{equation}
  \frac{1}{2}\left(\varrho^m-\frac{1}{\varrho^m}\right)  \leq|T_m(w) | \leq \frac{1}{2}\left(\varrho^m+\frac{1}{\varrho^m}\right).
\end{equation}
This shows that outside of the $[-1,1]$ interval the Chebyshev polynomials diverge exponentially as a function of $m$ (dashed lines in Fig.~\ref{fig:terms}).
In combination with the Bessel function, the terms get smaller with increasing $m$, ensuring the convergence of the expansion in Eq.~\eqref{eq:expansion}.

We provide an analytic upper bound for the value of the terms in the expansion.
For the Bessel function, the upper bound reads
\begin{equation}
    |J_m(t)| \leq \frac{1}{m!}\left(\frac{t}{2}\right)^m,
\end{equation}
where we used the asymptotic form for $m\gg t$ of the Bessel functions (that gives an upper bound).
From this formula it is easy to see how the Bessel functions decrease as a function of increasing $m$ more rapidly than the exponential increase of the Chebyshev polynomials.
We can also see that with increasing $t$ the maximum of the Bessel function moves to higher $m$, so the number of terms needed in the expansion to reach convergence also increases (see also Fig.~\ref{fig:terms}).

The numerical error of the expansion arises because of rounding errors in the floating point arithmetic.
We make the argument for real numbers, but for complex numbers it works similarly as they are just two copies of real numbers.
A real number $x\in\mathbb{R}$ is represented as
\begin{equation}
     x_\text{float} = s b^e,
\end{equation}
where $s\in\mathbb{Z}$ is the significand, $b$ is the base (usually 2 in most implementations), and $e$ is the exponent.
In usual cases when using double precision floats the significand is 53 bits.
This makes the largest value for the significand to be $s_\text{max}\approx \num{0.9e16}$.
This means that roughly the 17th digit (in a decimal system) is lost.
For an arbitrary number $x$ we can give an upper bound for this error (not taking into account errors accumulated during computation, simply the error arising from storing the number on a computer) as
\begin{equation}
|x-x_\text{float}| \leq \frac{x_\text{float}}{2^{53}} \equiv \varepsilon x_\text{float},
\end{equation}
where $\varepsilon=\num{1.11e-16}$ is the machine precision (for double precision floating point numbers).

We estimate the numerical error for the $m$-th term in the expansion as 
\begin{equation}
\begin{split}
    \Delta_m(\varrho,t) &\leq \varepsilon 4 |J_m(t)||T_m(w)| \leq \\ 
    &\leq 2 \varepsilon \left(\varrho^m+\varrho^{-m}\right)\frac{1}{m!}\left(\frac{t}{2}\right)^{m},
    \end{split}
\end{equation}
where the extra factor of $2$ comes from the relative errors of the Bessel functions and the Chebyshev polynomials adding up when multiplying the two numbers.
We assume that the precision of the Bessel function computed with floating point arithmetic
is $\varepsilon$ and the numerical errors coming from computational errors is smaller than the errors coming from the representation in the Chebyshev polynomials.
This is justified by the simple nature of the recursion of Chebyshev polynomials that uses only multiplications and additions.
We then take a larger upper bound that has a simpler form and does not significantly overestimate the error:
\begin{equation}
    \Delta_m(\varrho,t) < 4\varepsilon \frac{1}{m!}\left(\frac{t\varrho}{2}\right)^{m} \equiv \tilde\Delta_m(\varrho,t).
\end{equation}

The rounding error after adding $N$ numbers, each having $\varepsilon$ precision, can be computed as \cite{wilkinson1960}
\begin{equation}
    \tilde\Delta(\varrho,t) = \sum\limits_{m=0}^{N-1}\varepsilon^{(N)}_m \tilde\Delta_m(\varrho,t),
\end{equation}
where $\varepsilon_m^{(N)} \leq (N-1-m)$ for $m>0$ and $\varepsilon_0^{(N)} \leq N-2$.
As a simple upper bound for this total error we take
\begin{equation}
    \tilde\Delta < N \sum\limits_{m=0}^{\infty} \tilde\Delta_m = 4\varepsilon N \exp{\frac{t\varrho}{2}}.
\end{equation}

In order to estimate the number $N$ we use the condition
\begin{equation}
     \frac{4}{N!}\left(\frac{t\varrho}{2}\right)^{N} < \sum\limits_{m=0}^{\infty} \tilde\Delta_m  = 4\varepsilon \exp{\frac{t\varrho}{2}}, 
\end{equation}
which means that the sum is evaluated until the $N$-th term in the sum becomes smaller than the rounding error of the expansion.
Using the Stirling's approximation we approximate the factorial as $N! > (N/e)^N$ (here we took a simpler lower bound than the usual Stirling's approximation that contains a factor of $\sqrt{2\pi N}$).
This way the condition simplifies to
\begin{equation}
    N\log{\frac{t\varrho e}{2N}} < \log{\varepsilon+\log{N}+\frac{t\varrho}{2}}.
\end{equation}
For simplicity, we take a stricter condition
\begin{equation}
    N\log{\frac{t\varrho e}{2N}} < \log{\varepsilon+\frac{t\varrho}{2}},
\end{equation}
which we solve using the Lambert W function
\begin{equation}
    N > \frac{t\varrho e}{2}\exp{W\left(-2\frac{\log{\varepsilon}+\frac{t\varrho}{2}}{t\varrho e}\right)},
\end{equation}
where $W(x)\exp{W(x)}=x$ (we use the principal branch of the Lambert W function).
With this formula the total error becomes
\begin{equation}
\label{eq:errorfull}
    \tilde\Delta < 4\varepsilon \frac{t\varrho e}{2}\exp{W\left(-2\frac{\log{\varepsilon}+\frac{t\varrho}{2}}{t\varrho e}\right)} \exp{\frac{t\varrho}{2}}.
\end{equation}
For $t\varrho\to \infty$ this formula simplifies to
\begin{equation}
\label{eq:errorasymp}
    \tilde\Delta < 2\varepsilon t\varrho \exp{\frac{t\varrho}{2}},
\end{equation}
where we used $W(-1/e)=-1$.

\begin{figure}[t]
        \includegraphics[width=\columnwidth]{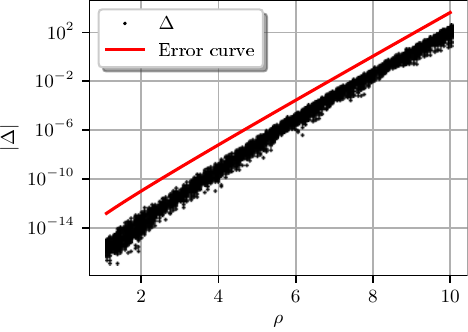} 
        \caption{\label{fig:errorcurve} The absolute difference ($\Delta$) between $\mathrm{e}^{-itz}$ computed using the standard exponential function and obtained using the Chebyshev series expansion, Eq.~\eqref{eq:expansion}, for different $\rho$ radius values of $z\in\mathbb{C}$. The order of the expansion is $m = 250$. The result is shown for $t=8$. The red curve shows the estimate for the rounding error computed using Eq.~\eqref{eq:errorasymp}.}
\end{figure}

The rounding error estimate in Eq.~\eqref{eq:errorasymp} is our first main result.
Figure~\ref{fig:errorcurve} shows this estimated rounding error together with the numerical results as a function of $\varrho$.
As we can see, the estimated error follows the same trend as the numerical data and gives a consistent upper bound.

\begin{figure}[t]
        \includegraphics[width=\columnwidth]{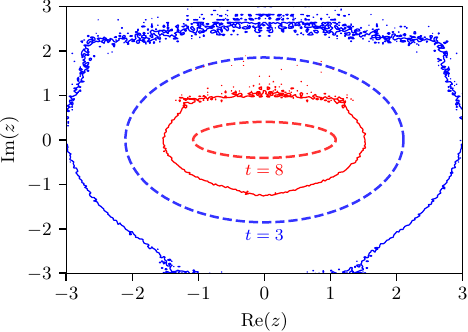} 
        \caption{\label{fig:ellipses_diff_t} Contours in the complex plane where the absolute difference $\Delta$ between $\mathrm{e}^{-itz}$ computed via the standard exponential and the Chebyshev expansion Eq.~\eqref{eq:expansion} satisfies $|\Delta| = 10^{-12}$, for times $t = 3,8$. The order of the expansion is $m = 250$.  The dashed lines show the Eq.~\eqref{eq:varrhomax} estimate for the contour and the solid contours show the numerical result.}
\end{figure}

The formula can also be inverted to give a safe upper radius $\varrho_\text{max}$ to use for any given time $t$ and absolute tolerance $\Delta_\text{max}$.
This constitutes our second main result:
\begin{equation}
\label{eq:varrhomax}
    \varrho_\text{max} = \frac{2}{t}W\left(\frac{\Delta_\text{max}}{4\varepsilon}\right).
\end{equation}
From this relation we can see that the longer the time step we want to compute, the smaller the radius we can use.
This is shown in Fig.~\ref{fig:ellipses_diff_t} for two different time steps, where the contours with $|\Delta|=10^{-12}$ are shown both numerically and from the analytic estimate.
Since we have a rigorous upper bound for the error estimate, the analytic ellipses are fully enclosed by the numerical contours, showcasing that this formula can be used to safely estimate the maximal radius that can be used during numerical computations.

\begin{figure}[t]
          \includegraphics[width=\columnwidth]{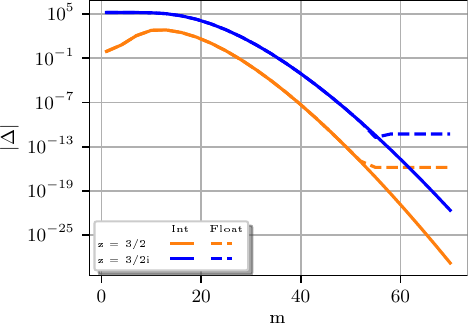}
        \caption{\label{fig:intfloat}The Chebyshev expansion of the complex exponential function \( e^{-i z t} \) is shown using integer -- denoted with Int -- and floating-point -- denoted with Float -- arithmetic using $t = 8$. The parameter \(\Delta\) represents the difference between the Chebyshev approximation from the exact exponential function, while \(m\) denotes the number of terms considered in the expansion.
        }
\end{figure}

In order to illustrate that this is only a numerical rounding error, we compare floating-point and integer arithmetic when computing Eq.~\eqref{eq:expansion} in Fig.~\ref{fig:intfloat}.
Using the integer representation is only possible with rational values for the real and imaginary parts of $z$.
Since the largest integer that we can represent exactly is significantly larger than the largest integer we can represent precisely as a float, the above explained numerical error will become relevant only at much larger values of $t$ or $\varrho$.

Figure~\ref{fig:intfloat} shows that the two representations yield similar approximations up to a specific order of the expansions, after which the float representation stops improving with increasing $m$, while the integer representation continues to improve.

\begin{figure}[t]
        \includegraphics[width=\columnwidth]{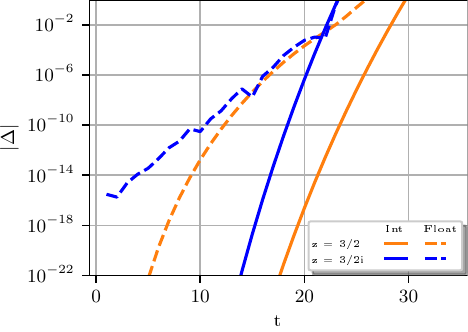} 
        \caption{\label{fig:Deltat}\ The time dependence of Chebyshev expansion of the complex exponential function \( e^{-i z t} \) is shown using integer -- denoted with Int -- and floating-point -- denoted with Float -- arithmetic. \(\Delta\) represents the difference between the Chebyshev approximation and the exact exponential function. The order in the expansion is $m = 100$.}

\end{figure}

Finally, the $t$ dependence of the expansion is shown in Fig.~\ref{fig:Deltat}.
Since a larger $t$ requires higher orders in the expansion, it leads to more numerical errors.
This appears as a strong monotonous increase in the error as a function of $t$.
For a fixed accuracy goal $|\Delta|$ there is a threshold value of $t_\text{max}$ above which the approximation becomes numerically bad.
We see that $t_\text{max}$ is significantly larger for the integer representation for the same $z$ value.

\section{Non-Hermitian time evolution in the Hatano-Nelson model}
\label{sec:expmat}

In this section we will generalize the expansion of the exponential function of a single complex number to a non-Hermitian matrix.
We then use this to compute the time evolution operator of non-Hermitian Hamiltonians and to compute the time evolution of an arbitrary initial state.
The method is completely general and can be used for any Hamiltonian given as a matrix.
To showcase the strengths and limitations of the method we will take the Hatano-Nelson (HN) model \cite{hatano1996} as an example where the time-evolution problem can be solved analytically.
The HN model is a one-dimensional (1D) tight-binding chain with non-reciprocal nearest neighbor hoppings
\begin{equation}
\label{eq:hatano}
\begin{split}
\hat{H} = &\gamma\sum\limits_{n=1}^{N-1} \bigg[(1+p)\ketbra{n}{n+1} + (1-p)\ketbra{n+1}{n} \bigg] \\
&+\alpha_{\text{BC}}\bigg[(1+p)\ketbra{N}{1}+(1-p)\ketbra{1}{N}\bigg],
\end{split}
\end{equation}
where $N$ is the number of sites in the chain, $\gamma$ is the energy scale of the hoppings, $p\in\mathbb{R}$, $|p|<1$ is the non-reciprocity in the hoppings (for $p=0$ we get a Hermitian 1D chain), $\ket{n}$ is the state localized on the $n$-th site, and $\alpha_{\text{BC}}=0$ for open boundary condition (OBC) and $\alpha_{\text{BC}}=1$ for periodic boundary condition (PBC).
In the $\ket{n}$ basis the components of the Hamiltonian matrix are $H_{mn}=\bra{m}\hat H\ket{n}$.

\subsection{Chebyshev expansion of the time evolution operator}
Since the Chebyshev expansion of the exponential function works on the entire complex plane, its generalization to non-Hermitian matrices is straightforward
\begin{equation}
\label{eq:expansion_mat}
        \mathrm{e}^{-itH} = J_0(t)+2\sum\limits_{m=1}^\infty (-i)^mJ_m(t)T_m(H),
\end{equation}
where $H$ is an arbitrary square matrix and the Chebyshev polynomials are given by the same recursive relation
\begin{equation}
\begin{split}
            T_0(H)&=1,\\
        T_1(H)&=H, \\
        T_{m+1}(H)&=2zT_m(H)-T_{m-1}(H).
\end{split}
\end{equation}

This is true because the expansion works for all eigenvalues of any complex matrix, which means it must also work for the matrix itself. This statement is straightforward for matrices that can be diagonalized, but it also applies to matrices that are non-diagonalizable.
Every matrix can be brought to a Jordan normal form which is a block diagonal matrix composed of Jordan blocks
\begin{align}
H &= P \left[\bigoplus_{a=1}^n J_a \right] P^{-1}, &
J_a&= 
    \begin{pmatrix}
\varepsilon_a & 1       & 0       & \cdots  & 0 \\
0       & \varepsilon_a & 1       & \ddots  & \vdots \\
0       & 0       & \varepsilon_a & \ddots  & 0 \\
\vdots  & \ddots  & \ddots  & \ddots  & 1 \\
0       & \cdots  & 0       & 0       & \varepsilon_a
\end{pmatrix},
\end{align}
where $\varepsilon_a$ are the eigenvalues of the $H$ matrix and $P$ is an invertible matrix.
An f(H) analytic function can then be evaluated as 
\begin{subequations}
\begin{align}
f(H) &= P \left[\bigoplus_{a=1}^n f(J_a) \right] P^{-1},\\
    f(J_a) &=
\begin{pmatrix}
f(\varepsilon_a) & f'(\varepsilon_a) & \tfrac{f''(\varepsilon_a)}{2!} & \cdots & \tfrac{f^{(n-1)}(\varepsilon_a)}{(n-1)!} \\
0 & f(\varepsilon_a) & f'(\varepsilon_a) & \ddots & \vdots \\
0 & 0 & f(\varepsilon_a) & \ddots & \tfrac{f''(\varepsilon_a)}{2!} \\
\vdots & \ddots & \ddots & \ddots & f'(\varepsilon_a) \\
0 & \cdots & 0 & 0 & f(\varepsilon_a)
\end{pmatrix}.
\end{align}
\end{subequations}
In our case $f(\varepsilon_a)=\mathrm{e}^{it\varepsilon_a}$ is analytic and can be expressed using the Chebyshev expansion for any $\varepsilon_a\in\mathbb{C}$. This means that the Chebyshev expansion of the exponential function works for every Jordan block and thus can be applied to any square matrix.

\subsection{Analytic time evolution}

The time-evolution problem can be solved formally using the eigenvalues and eigenstates of the Hamiltonian.
\begin{equation}
    \hat{H}\ket{\psi_a}=\varepsilon_a\ket{\psi_a}.
\end{equation}
The time evolution of a state $\ket{\phi}$ is expressed as
\begin{equation}
\label{eq:basisevolve}
    \ket{\phi(t)} = \sum\limits_a \mathrm{e}^{-i\varepsilon_a t} \braket{\psi_a}{\phi} \ket{\psi_a}.
\end{equation}
In the following we give the analytic solutions to the eigenvalue problem of the Hatano-Nelson model under PBC and OBC.

\subsubsection{Periodic boundary condition}

The PBC Hatano-Nelson model [$\alpha_{\text{BC}}=1$ in Eq.~\eqref{eq:hatano}] in the $\ket{n}$ basis is a circulant matrix of the following form
\begin{equation}
\label{eq:PBC}
H = \gamma\begin{pmatrix}
0 & 1 + p & 0 & \cdots & 0 & 1 - p \\
1 - p & 0 & 1 + p & \cdots & 0 & 0 \\
0 & 1 - p & 0 & \cdots & 0 & \vdots \\
\vdots & \vdots & \vdots & \ddots & \vdots & 0 \\
0 & 0 & 0 & \cdots & 0 & 1+p \\
1 + p & 0 & 0 & \cdots & 1 - p & 0
\end{pmatrix}.
\end{equation}

The eigenvalue problem $\sum_m H_{nm}\psi_m^{(a)} = \varepsilon_a \psi_m^{(a)}$ of circulant matrices can be solved via Fourier transform \cite{gray2006} and the eigenvalues $\varepsilon$ and right eigenvectors $\psi$ of Eq.~\eqref{eq:PBC} are expressed as

\begin{subequations}
\label{eq:pbcspectrum}
\begin{align}
    \varepsilon_a &= \gamma(1 + p)e^{-i \frac{2\pi a}{N}} + \gamma(1 - p)e^{i\frac{2\pi a}{N}},\\
    \psi_n^{(a)} &= \frac{1}{\sqrt{N}} e^{-i n \frac{2\pi a}{N}},
\end{align}
\end{subequations}
where $a,n=1,\cdots,N$.

\subsubsection{Open boundary condition}

The OBC Hatano-Nelson model [$\alpha_{\text{BC}}=0$ in Eq.~\eqref{eq:hatano}] in the $\ket{n}$ basis is a tridiagonal Toeplitz matrix of the following form
\begin{equation}
\label{eq:PBC}
H = \gamma\begin{pmatrix}
0 & 1 + p & 0 & \cdots & 0 & 0  \\
1 - p & 0 & 1 + p & \cdots & 0 & 0  \\
0 & 1 - p & 0 & \cdots & 0 & 0  \\
\vdots & \vdots & \vdots & \ddots & \vdots & \vdots  \\
0 & 0 & 0 & \cdots & \ddots &1+p  \\
0 & 0 & 0 & \cdots & 1-p &0  \\
\end{pmatrix}.
\end{equation}

The eigenvalues and right eigenvectors of this Hamiltonian are given by \cite{noschese2013}:
\begin{subequations}
\label{eq:obcspectrum}
\begin{align}
    \varepsilon_a &= 2 \gamma\sqrt{(1 - p)(1 + p)} \cos\left(\frac{a \pi}{N + 1}\right),\\
    \psi_{n}^{(a)} &= \left(\frac{1 - p}{1 + p}\right)^{\frac{n}{2}} \sin\left(\frac{a n \pi}{N + 1}\right),
\end{align}
\end{subequations}
where $a,n=1,\cdots,N$.

\subsection{Numerical time evolution}

Here, we use the Chebyshev expansion to calculate the time evolution of $|\psi(0)\rangle$
\begin{equation}
        |\phi(t)\rangle=\mathrm{e}^{-itH} |\phi(0)\rangle.
\end{equation}
Numerically, it is faster to compute the effect of the time evolution operator on the initial state, than first computing the time evolution operator and then acting with it on the initial state.
Using the Eq.~\eqref{eq:expansion_mat} expansion and applying it to the initial state, we directly compute the time evolved state as
\begin{equation}
\label{eq:chebwave}
        |\psi(t)\rangle = J_0(t)|\psi(0)\rangle+2\sum\limits_{m=1}^\infty (-i)^mJ_m(t)T_m(H)|\psi(0)\rangle,
\end{equation} where
\begin{align}
\begin{split}
        T_0(H)|\psi\rangle&=|\psi\rangle,\\
        T_1(H)|\psi\rangle&=H|\psi\rangle, \\
        T_{m+1}(H)|\psi\rangle&=\left[2HT_m(H)-T_{m-1}(H)\right]|\psi\rangle.
        \end{split}
\end{align}
The advantage of this is that only matrix-vector multiplications are needed instead of matrix-matrix multiplications.

We compute the time evolution for a wave packet using the Chebyshev expansion and compare the result with the analytic solution and take the norm of differences as error.
We calculate the time evolution for a fixed $T_{max}$ total time splitting it into steps of $\Delta t$. In order to keep the wave packets normalized we rescale them at each time step.
To optimize the algorithm, instead of using a fixed $m$ expansion we use an adaptive approach where the series is stopped once the terms become smaller than $10^{-14}$ for $5$ consecutive iterations. This ensures that in cases where numerical convergence is achieved with fewer terms we avoid unnecessary computation.

\begin{figure}[tb]
        \includegraphics[width=\columnwidth]{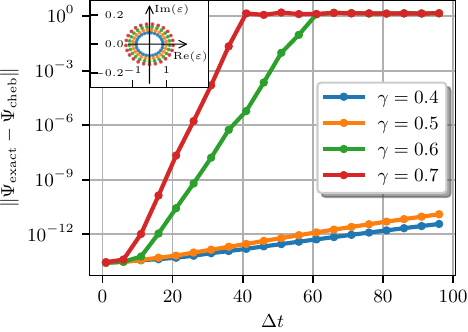} 
        \caption{\label{fig:HNPBC}The norm of the difference between the exact time-evolved wave packet and the wave packet obtained using the Chebyshev expansion, Eq.~\eqref{eq:chebwave}, as a function of the time step $\Delta t$ using the Hatano-Nelson model, Eq.~\eqref{eq:hatano}. The simulation parameters used are $T_{max} = \Delta t$ maximum time, for the system with $N=100$ system size, $\gamma=0.4,0.5,0.6,0.7$ hopping energy, and $p=0.1$ non-reciprocity under periodic boundary condition. The initial wavefunction is a Gaussian wave packet with $k=\pi/2$ momentum centered on the middle of the chain, with width $\sigma=10$. The inset shows the different spectra on the complex energy plane. }The parameters are such that for the two smaller (larger) values of $\gamma$ the real part of the spectrum is inside (outside) the $[-1,1]$ interval.
\end{figure}

\begin{figure}[tbh]
        \includegraphics[width=\columnwidth]{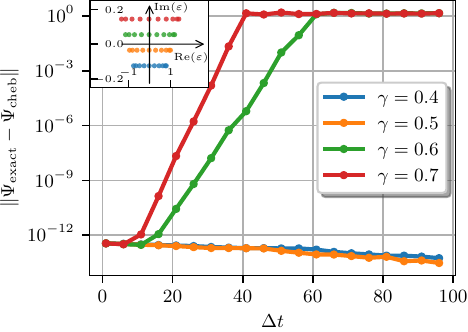} 
        \caption{\label{fig:HNOBC} Same simulations with the same parameters as the ones in Fig.~\ref{fig:HNPBC}, but under open boundary conditions. The inset shows the spectrum which is completely real in this case (the different spectra are shifted for better visibility).}
\end{figure}

First we look at a single time step of $\Delta t$.
The results as function of $\Delta t$ for the PBC and OBC Hatano-Nelson model are shown in Figs.~\ref{fig:HNPBC} and \ref{fig:HNOBC}.

In the PBC case, by Eqs.~\eqref{eq:pbcspectrum}, the spectrum lies on an ellipse in the complex plane. Increasing $\gamma$ increases the size of the ellipse and thus decreases the largest $\Delta t$ that still gives an accurate result.
In the OBC case, by Eqs.~\eqref{eq:obcspectrum}, the spectrum is always real. As long as it is in the $[-1,1]$ interval the expansion works very well.
For the $\gamma$ values where the spectrum goes beyond the $[-1,1]$ interval the expansion works well only for smaller $\Delta t$ steps.

The two cases work similarly well, with the OBC working better for the ranges of $\gamma$ where the real part of the spectrum lies within the $[-1,1]$ interval due to the fully real spectrum.
Interestingly, the expansion works quite well even for the PBC in this range, even though the spectrum is not fully real. 

Comparing the numerical results for larger systems and larger non-reciprocity becomes difficult in the case of OBC. This is because the representation of the analytic results in the computer becomes imprecise due to the exponentially localized eigenstates. When computing the overlaps $\bra{\psi_a}\ket{\phi}$ in Eq.~\eqref{eq:basisevolve} the small numbers in the exponential tail cause numerical errors. This shows the limitations of the exact diagonalization or of the analytic approach, which the Chebyshev expansion method does not have.

\begin{figure}[tbh]
        \includegraphics[width=\columnwidth]{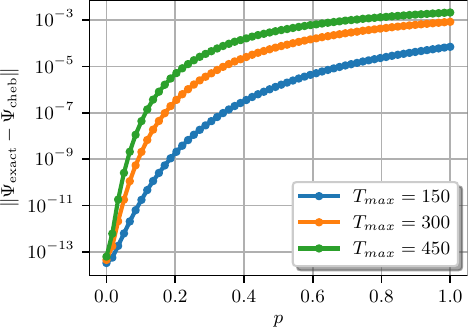} 
        \caption{\label{fig:HNPBCp}The norm of the difference between the exact time-evolved wave packet and the wave packet obtained using the Chebyshev expansion Eq.~\eqref{eq:chebwave} as a function of the non-reciprocity $p$ using the Hatano-Nelson model, Eq.~\eqref{eq:hatano}. The simulation parameters used are $\Delta t=1$ timestep, $T_{max} = 150,300,450$ maximum time, for a system size $N=100$ and for $\gamma=1$, under periodic boundary condition. The initial wavefunction is a Gaussian wave packet with $k=\pi/2$ momentum centered at the middle of the chain, with width $\sigma=10$.}
\end{figure}

In the PBC case we do not have this limitation as there the eigenstates are ordinary plane waves. Figure~\ref{fig:HNPBCp} shows different simulations as a function of the non-reciprocity $p$ in the Hatano-Nelson model for PBC.

The increasing value of $p$ increases the $\varrho$ radius of the Bernstein ellipse enclosing the spectrum of the Eq.~\eqref{eq:hatano} Hamiltonian.
In the PBC case the spectrum is always complex, thus the error increases monotonously with $p$ until it reaches the maximum value (enforced by the normalization of the wavefunctions).
The figure also shows that longer simulations with larger $T_{max}$ are less accurate.

\begin{figure}[tb]
        \includegraphics[width=\columnwidth]{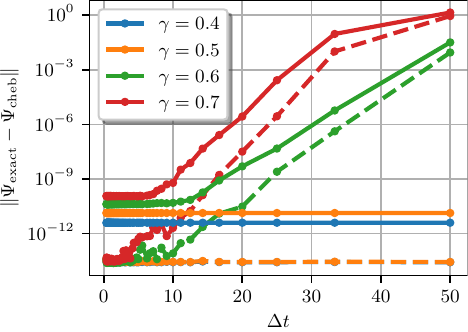} 
        \caption{\label{fig:HNPBCdt}The norm of the difference between the exact time-evolved wave packet and the wave packet obtained using the Chebyshev expansion Eq.~\eqref{eq:chebwave} as a function of a single time step $\Delta t$ using the Hatano-Nelson model \eqref{eq:hatano}. The simulation parameters used are $T_{max} = 100$ maximum time, for a system size $N=100$, $\gamma=0.4,0.5,0.6,0.7$, and $p=0.1$ (solid lines) $p=0$ (dashed lines) non-reciprocity under periodic boundary condition. The initial wavefunction is a Gaussian wave packet with $k=\pi/2$ momentum centered at the middle of the chain, with width $\sigma=10$. The parameters are such that for the two smaller (larger) values of $\gamma$ the real part of the spectrum is inside (outside) the $[-1,1]$ interval.}
\end{figure}

We finally show a simulation where the final time $T_{max}$ is kept constant and we vary the number of steps in Fig.~\ref{fig:HNPBCdt}. This is how realistically a simulation would go.

As can be seen, it is possible to obtain numerically very precise results with the right choice of time steps. If we compare the results to that of the Hermitian $p=0$ case we see that we get comparably good results, demonstrating how the method not only applies for Hermitian Hamiltonians and not only within the range $[-1,1]$.

\section{Conclusions}
\label{sec:conc}

We showed that the Chebyshev expansion method to compute the time evolution of a quantum state can be extended from Hermitian systems with a spectrum restricted to the $[-1,1]$ interval to non-Hermitian systems with arbitrary complex spectra.

Numerical rounding errors cause the expansion to break down for too large time steps.
This means that for numerically accurate results the time step must be chosen based on the spectrum of the Hamiltonian.
The larger the $\varrho$ radius of the Bernstein ellipse containing the whole spectrum, the smaller $\Delta t$ step must be taken.

Using the Hatano-Nelson model, we demonstrated that the analytic estimates provided in Eqs.~\eqref{eq:errorasymp} and \eqref{eq:varrhomax} can be used to select appropriate simulation parameters such that the numerical errors remain below a desired value.

\textit{\textcolor{blue}{Acknowledgments}} -- This work was supported by the Deutsche Forschungsgemeinschaft (DFG, German Research Foundation) under Germany’s Excellence Strategy through the W\"{u}rzburg-Dresden Cluster of Excellence on Complexity and Topology in Quantum Matter – ct.qmat (EXC 2147, 390858490 and 392019). V.K. was funded by the European Union.
D.~V. was supported by the National Research, Development and Innovation Office of Hungary under OTKA grant no. FK 146499, and
the János Bolyai Research Scholarship of the Hungarian Academy of Sciences.
O.~L. was supported by the Ministry of Culture and Innovation and the National Research,
Development and Innovation Office within the Quantum Information National Laboratory of Hungary
(Grant No. 2022-2.1.1-NL-2022-00004), National Research, Development and Innovation Office (NKFIH) through Grant Nos. K134437 as well as projects KKP133827 and	K142179.
This project is supported by the TRILMAX Horizon Europe consortium (Grant No. 101159646).

\bibliography{references}

\appendix

\section{Numerical Implementation}
\label{sec:implementation}
Our implementation for the Chebyshev expansion is available at Ref.~\cite{zenodo}.
For obtaining the standard exponential function using floating-point arithmetic we used the built-in exponential function of the \texttt{NumPy} package.
For the integer arithmetic we used the \texttt{SymPy} package.
\end{document}